# "Fractal Expression" in Chinese Calligraphy

Yuelin Li*

We show that from historical record and mathematic analysis, "fractal expressions" may have been a conscious pursuit at least one thousand years ago as an element of beauty in ancient Chinese calligraphy.

The finding of fractal dimensions [1] in drip paintings of Jackson Pollock [2], an enigmatic American abstract expressionist, has sparked a sensation among scholars in different disciplines. It was even marveled that "Pollock was honing his ability to generate fractals a full quarter century before fractal geometry was formally described" [3].

Perplexing as this is, Pollock left scarcely any remarks about his art (among them the famous "make energy visible" and "no chaos"). The meaning and expression of his art remain a topic of debate, and even the measured fractal is being questioned as a design [4]. In any case, even Pollock truly grasped the "fractal expression," in fact, the conscious pursuit of fractal beauty in art may be dated much earlier.

One example is Chinese calligraphy. Calligraphy became a leisure pursuit among scholars and government officials about 2000 years ago in China, when Chinese scripts transcended from a pictograph system into a modern script. Still enjoyed by thousands in eastern Asia,

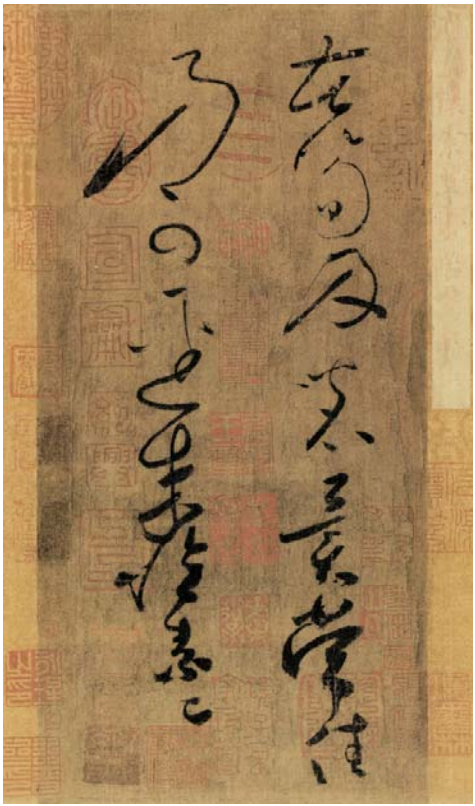

Fig. 1: Huai Su, Bitter bamboo shoot letter, ink on silk, 12 cm × 25 cm, collection of Shanghai Museum, Shanghai. It says "Bitter bamboo shoots and tea? Excellent! Just rush them, presented by Hui Su."

Chinese calligraphy employs a soft brush to write on paper, fabric, and other flat surfaces. Noticeably, traditional praise and interpretation of the calligraphy's aesthetic beauty always alludes to the variety in nature's scenes, yet its visual never goes beyond a monochromatic labyrinth of brush traces [5]. No example can illustrate this better than a highly decorated dialogue recorded in 722 CE. It occurred between two of the most legendary figures in Chinese calligraphy history: Huai Su, a maniacal Buddhist monk devoted to cursive script, and Yan Zhenqing, a Confucian government official and renowned calligrapher [6],

> "……Zhenqing asked: "Do you have your own inspiration?" Su answered: "I often marvel at the spectacular summer clouds and imitate them …… I also find the cracks in a wall very natural." Zhenqing asked: "How about water stains of a leaking house?" Su rose, grabbed Yan's hands, and exclaimed: "I get it!"

This conversation has virtually defined the aesthetic standard of the Chinese calligraphy thereafter and the "house leaking stains" and "wall cracks" became a golden measure of the skill of a calligrapher and the quality of his work. Ironically, a consensus on the interpretation of Huai Su's "eureka" moment and the true aesthetic meaning of the three visuals has never been reached.

The conversation alludes to, in quick succession, three natural observations: summer clouds, the cracks in a wall, and water stains. On the surface, they are formed by different physical processes: turbulent diffusion for clouds, bound breaking for fractures, and self-organizing flow (similar to a meandering river) and wetting of a surface for the water stain. The only visual characteristic they have in common is randomness. However, as revealed by modern science, their hidden geometry is the same: the fractal [1]. Perhaps, the artists sensed an intricate form of order beneath the apparent randomness of the clouds, cracks, and water stains that they regarded as nature's visual secret and from which they derived their brush expression.

Were they successful in capturing the hidden order and grasping the "fractal expression" in their works? Fractal analysis [7] of one of the authentic pieces by Huai Su (Figure 1), a letter begging a friend to rush some bamboo shoots and tea, reveals two distinctive fractal dimensions: $D_s$=1.74 ±0.01 for a smaller range of 0.01 cm < $L$ <0.1 cm, roughly the size of the brush strokes and smaller; and $D_L$=1.41 ±0.02 for a larger range 0.1 cm < $L$ < 12 cm, a range that covers the character size and the dimension of the piece [8]. Figure 2 demonstrates the visual similarity of the work at different scales.

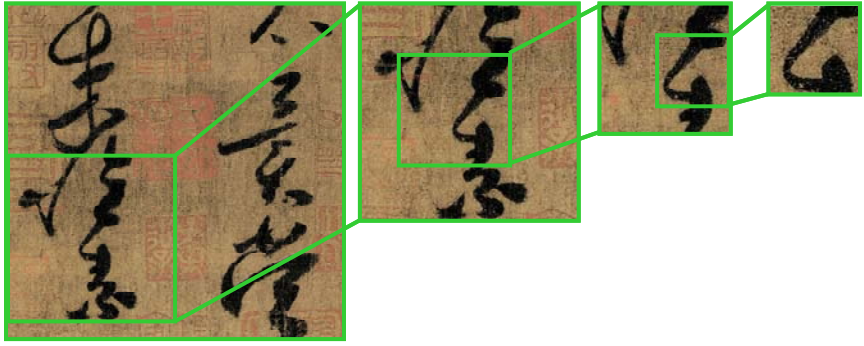

Figure 2: Details of Figure 1 showing a degree of visual self similarity and complexity.

Two aspects should be noted for this observation. First, the measured fractal is not to be taken as mathematically rigid due to the non-mathematical nature of human motion control and visual feedback of human physical capability, a point largely ignored in a recent Pollock fractal debate [2, 4]. The box counting technique, on the other hand, does not capture the particularity of a fractal pattern (such as the effect of strange attractors), thus the data can be open to different interpretations.

In the light of the above historic record, the fractal can be attributed to the artist's pursuit of the hidden order of the fractal. The fractal for the smaller dimension is the variation of the brush traces (size and texture) by controlling the wetting of the medium via brush motion: the orientation, speed, pressure, and the quantity of the ink contained in the brush, etc. The wear due to aging of the piece (which by itself is fractal and adds to the aesthetic value) may also have contributed. The larger-scale fractal is mainly the result of the artist's technique in combining the chaotic, nonlinear hand motion during writing [9] and the complex construction of the Chinese script. Importantly, while Pollock relied on the integrated effect of the overlaid paint layers to achieve the apparent fractal in his paintings, Huai Su's calligraphy is a strict time-sequence composition of the richly varying traces of his brush.

It remains to be understood how artists, separated by more than a millennium and living in drastically different cultural and technological environments, arrived at similar expressions of abstract imagery. Is the fractal, like symmetry [10], one of the synaptic rules of the figural primitives of our perceptual grammar speculated by Ramachandran [11]?

___________
* Yuelin Li is a physicist at Argonne National Laboratory, Argonne, IL 60439, ylli@aps.anl.gov